# Signal and Noise Analysis for Chiral Structured Illumination Microscopy

SHIANG-YU HUANG[1*], ANKIT KUMAR SINGH[1], AND JER-SHING HUANG[1,2,3,4*]

[1] *Leibniz Institute of Photonic Technology, Albert-Einstein Straße 9, 07745 Jena, Germany*
[2] *Abbe Center of Photonics, Friedrich-Schiller University Jena, Jena, Germany*
[3] *Research Center for Applied Sciences, Academia Sinica, 128 Sec. 2, Academia Road, Nankang District, 11529 Taipei, Taiwan*
[4] *Department of Electrophysics, National Chiao Tung University, 1001 University Road, 30010 Hsinchu, Taiwan*
*\*shiang-yu.huang@leibniz-jena.de; jer-shing.huang@leibniz-jena.de*

**Abstract:** Recently, chiral structured illumination microscopy has been proposed to image fluorescent chiral domains at sub-wavelength resolution. Chiral structured illumination microscopy is based on the combination of structured illumination microscopy, fluorescence-detected circular dichroism, and optical chirality engineering. Since circular dichroism of natural chiral molecules is typically weak, the differential fluorescence is also weak and can be easily buried by the noise, hampering the fidelity of the reconstructed images. In this work, we systematically study the impact of the noise on the quality and resolution of chiral domain images obtained by chiral SIM. We analytically describe the signal-to-noise ratio of the reconstructed chiral SIM image in the Fourier domain and verify our theoretical calculations with numerical demonstrations. Accordingly, we discuss the feasibility of chiral SIM in different experimental scenarios and propose possible strategies to enhance the signal-to-noise ratio for samples with weak circular dichroism.

## 1. Introduction

Chirality, which is the lack of mirror symmetry in three-dimensional conformation, is of crucial importance in various research fields as it governs the optical properties and functionalities of matters. While traditional spectroscopic methods, such as circular dichroism (CD) spectroscopy, have brought access to the molecular chirality of different targets [1,2], the collective signals measured by those spectroscopic methods are unable to provide spatial features of local chiral domains. In this regard, various imaging methodologies, such as direct CD imaging [3-7], nonlinear CD mapping [8-12], photothermal CD microscope [13-16], and CD scanning near-field optical microscope [17,18] have been proposed and demonstrated to offer graphic inspection and address the regional variations of chirality. However, limitations such as weak CD contrast, low image speed and diffraction-limited resolution still restrain the capability of the aforementioned chiral image methods.

Recently, we proposed a new chiral imaging method, namely chiral structured illumination microscopy (chiral SIM), to image chiral domains at sub-wavelength resolution [19]. In chiral SIM, the illumination with structured optical chirality (OC) patterns allows for imaging fluorescent chiral domains with superior resolving power. Moreover, chiral SIM is based on wide-field illumination and does not need to scan across the sample. Therefore, chiral SIM can be performed at a relatively high acquisition speed and is potentially useful for *in vivo* imaging of chiral targets at super-resolution. Despite these advantages, chiral SIM has not been experimentally demonstrated until now. One major challenge towards the experimental realization is that, for natural chiral molecules, the fluorescence signal modulation can be too small to overcome the noise that occurs during the image acquisition. This greatly limits the fidelity of the final reconstructed images. For chiral SIM to function correctly, the CD-dependent fluorescence signals, i.e., fluorescence-detected circular dichroism (FDCD), recorded in chiral SIM should not be buried by the noise. This is an important prerequisite for the subsequent image reconstruction in chiral SIM. To address this issue, we previously discussed the feasibility of chiral SIM by examining the modulation-depth-to-noise-ratio (MDNR), that is the ratio between the amplitude of the modulated fluorescence intensity and the noise [19]. Although MDNR provides an overview of the noise effects on the reconstructed chiral SIM image and illustrates what can be applied in order to enhance the image quality, it was based on the assumption of uniform sample and chirality distribution. For samples with complex chirality distribution, a comprehensive study of the chiral SIM image quality in the presence of noise is necessary to obtain optimal imaging conditions and inspect the applicability of versatile chiral samples to the proposed chiral SIM method.

In this work, we elucidate the processes of image acquisition and image reconstruction in chiral SIM and analytically derive the signal-to-noise ratio (SNR) of the reconstructed chiral SIM image as a function of spatial frequency. As a measure usually applied in noise analysis, SNR compares the expected signals with noise and is useful as a metric of image fidelity. Since the spatial features of complex samples can be Fourier decomposed into respective spatial frequencies, analysis of SNR in the Fourier space provides a better insight into the influence of the noise than that in the real space. We present numerical demonstrations of chiral SIM by applying a synthetic sample of the randomly distributed chiral filaments and discuss the capability of chiral SIM in different experimental conditions.

## 2. Principle of Chiral SIM

### 2.1 Image acquisition in chiral SIM

As a fluorescence microscopy method, chiral SIM detects the difference in fluorescence that depends on the CD of the molecules. This implies that the fluorophores need to obey the following criteria of FDCD. First, the fluorophores should be either chiral or attached to the chiral domains. Secondly, the rotatory Brownian motion should fully randomize the orientation of the fluorophore during their excited state lifetime [20-23]. With the criteria of FDCD being satisfied, the CD-dependent fluorescence emitted from chiral molecules within the exposure time $\Delta t$ is

$$F(\mathbf{r}) = \Phi_q \beta_d \eta \, \Delta t \, A(\mathbf{r}), \tag{1}$$

where $\Phi_q$ and $\beta_d$ are the quantum yield of the fluorophore and the detection efficiency of the optical system, respectively, and $\eta$ is the number of fluorophores within a voxel. $A(\mathbf{r}) = \frac{\varepsilon}{2}\left[\omega U_e(\mathbf{r})\alpha''(\mathbf{r}) - C(\mathbf{r})G''(\mathbf{r})\right]$ denotes the molecular absorption, where $C(\mathbf{r}) = -\frac{\varepsilon\omega}{2}\mathrm{Im}\left[\mathbf{E}^*(\mathbf{r})\cdot\mathbf{B}(\mathbf{r})\right]$ is the OC distribution of the excitation field and $U_e(\mathbf{r}) = \frac{\varepsilon}{4}\left|\mathbf{E}(\mathbf{r})\right|^2$ the time-averaged electric energy density distribution. $\varepsilon$ is the permittivity of the environment and $\omega$ denotes the angular frequency. $\mathbf{E}(\mathbf{r})$ and $\mathbf{B}(\mathbf{r})$ are time-harmonic complex local electric and magnetic fields, respectively [24]. Consider the point spread function (PSF) of the imaging system $h(\mathbf{r})$, the acquired CD-dependent fluorescence image is

$$M(\mathbf{r}) = \frac{2\Phi_q \beta_d \eta \, \Delta t}{\varepsilon}\left[\omega U_e(\mathbf{r})\alpha''(\mathbf{r}) - C(\mathbf{r})G''(\mathbf{r})\right] \otimes h(\mathbf{r}), \tag{2}$$

where $\otimes$ denotes the convolution operation. On the right-hand side of Eq. (2), the first term is attributed to the electric dipole absorption, which is usually dominant during the absorption process, and the second term is contributed by the chirality-sensitive absorption. In chiral SIM,

$C(\mathbf{r})$ of the illumination is periodically structured to externally modulate the chirality-dependent fluorescence. This leads to the generation of the moiré patterns on the fluorescent chiral domains, as illustrated in the right panel of Fig. 1. The fine details of the chiral domains that correspond to the high spatial frequencies can be thus down-modulated and captured by a diffraction-limited imaging system. Since the chiral response is accompanied by the signals from electric dipole absorption, as described by Eq. (2), the illumination fields should be smartly engineered to enable the separation of these two types of signal. One possible way is to construct a uniformly-distributed $U_e(\mathbf{r})$ (i.e., $U_e(\mathbf{r}) = $ const.) such that the portion of electric dipole absorption in the fluorescence image is unmodulated. In this sense, the signal from the electric dipole absorption can be removed either by subtraction in the real space or by allying a DC filter to in the Fourier space before image reconstruction. Alternatively, $U_e(\mathbf{r})$ of the illumination can also be designed to have complex spatial distribution. However, $U_e(\mathbf{r})$ should stay the same when the handedness of the distribution of $C(\mathbf{r})$ is flipped between right and left. In this way, the signals contributed by electric dipole absorption can be eliminated by the subtraction of two conjugate images in the real space [25]. Detailed discussion on the criteria of the illumination is provided in Appendix 6.1.

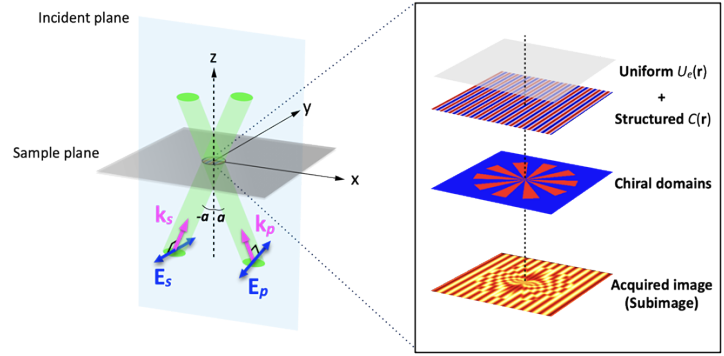

Fig. 1. Schematic of the illumination scheme and image acquisition in chiral SIM used in this work. Left panel: The illumination scheme using far-field optics. $\mathbf{E}_s$, $\mathbf{E}_p$, $\mathbf{k}_s$ and $\mathbf{k}_p$ indicate the electric field vectors and the wavevector of the $s$- and $p$-polarized light, respectively. $a$ denotes the incident angle of the linearly polarized light with respect to the surface normal of $xy$-plane. Right panel: Illustration of the subimages acquisition. The contrast of the moiré pattern in the subimage is enhanced for the clarity of presentation.

In our previous work, we have presented several illumination schemes that fulfill these requirements, such as using the interference of two circularly polarized or two linearly polarized light beams [26]. Here, we only focus on the scenario, where the illumination fields are formed by the superposition of counter-propagating $s$- and $p$-polarized light of equal amplitude, as shown in the left panel of Fig. 1. The resulting $C(\mathbf{r})$ and $U_e(\mathbf{r})$ generated by this illumination scheme can be expressed as [19]

$$C(\mathbf{r}) = C_0 \cos(\mathbf{k}_C \cdot \mathbf{r} + \varphi), \qquad (3)$$

$$U_e(\mathbf{r}) = \frac{\varepsilon}{4} E_0^2, \qquad (4)$$

where $E_0$ is the electric field amplitude, $C_0 = \dfrac{\varepsilon\omega}{2c}E_0^2\cos^2 a = C_{\mathrm{CPL}}\cos^2 a$ is the amplitude of the cosinusoidal OC pattern that relates to the OC of circularly polarized light (CPL) $C_{\mathrm{CPL}}$. $\mathbf{k}_C$ is the wavevector that indicates the orientation of the OC patterns and $|\mathbf{k}_C| = 2nk_0\sin a$ determines the spatial frequency of the structured patterns, where $n$ indicates the refractive index of the environment. $\varphi = \dfrac{\pi}{2} - (\varphi_s - \varphi_p)$ is the phase of the OC patterns and relates to the initial phase of the $s$-polarized light $\varphi_s$ and $p$-polarized light $\varphi_p$. $k_0$ is the wavevector of the excitation light in free space and $a$ is the incident angle of $s$- and $p$-polarized light relative to the surface normal of $xy$-plane, as indicated in Fig. 1. In the simulations, the power of the superimposed light beams has been normalized to that of a single CPL source for a fair comparison. According to Eq. (4), $U_e(\mathbf{r})$ is spatially invariant so that we can omit the spatial coordinate, namely $U_e(\mathbf{r}) = U_e$. With this illumination condition, the recorded subimage can be expressed as

$$M(\mathbf{r}) = \dfrac{2\Phi_q\beta_d\eta\Delta t}{\varepsilon}\left[\omega U_e \alpha''(\mathbf{r}) - C_0\cos(\mathbf{k}_C\cdot\mathbf{r}+\varphi)G''(\mathbf{r})\right]\otimes h(\mathbf{r}). \qquad (5)$$

Here, because $U_e$ is constant, the signals from the electric dipole absorption can be easily excluded in the Fourier analysis, as will be illustrated in the following section.

## 2.2 Image reconstruction in chiral SIM

Similar to the conventional SIM method, the image reconstruction of chiral SIM is performed in the Fourier space. Starting from Eq. (5), the Fourier transform of the acquired subimages is performed as

$$\begin{aligned}\tilde{M}(\mathbf{k}) &= \dfrac{2\Phi_q\beta_d\eta\Delta t}{\varepsilon}\left[\omega U_e \tilde{\alpha}''(\mathbf{k}) - \tilde{C}(\mathbf{k})\otimes\tilde{G}''(\mathbf{k})\right]\tilde{h}(\mathbf{k}) \\ &= \dfrac{2\Phi_q\beta_d\eta\Delta t}{\varepsilon}\left[\omega U_e \tilde{\alpha}''(\mathbf{k}) - C_0 e^{i\varphi}\tilde{G}''(\mathbf{k}-\mathbf{k}_C) - C_0 e^{-i\varphi}\tilde{G}''(\mathbf{k}+\mathbf{k}_C)\right]\tilde{h}(\mathbf{k}),\end{aligned} \qquad (6)$$

with a cosinusoidal form for $C(\mathbf{r})$ mentioned in Eq. (3). Here, the tilde ~ denotes the Fourier transform of the corresponding function, $\mathbf{k}$ is the coordinate vector that indicates the spatial frequency and direction in Fourier space, and $\tilde{h}(\mathbf{k})$ is the optical transfer function (OTF), which is the Fourier transform of the PSF. As shown in Eq. (6), three components that contain the signals resulting from the electric dipole absorption and CD responses are mixed in $\tilde{M}(\mathbf{k})$ simultaneously. For clarity, we can rewrite Eq. (6) as

$$\tilde{M}(\mathbf{k}) = -\dfrac{2\Phi_q\beta_d\eta\Delta t}{\varepsilon}\sum_{l=-1}^{1}\tilde{\Omega}_l(\mathbf{k}-l\mathbf{k}_C)e^{il\varphi}, \qquad (7)$$

where $\tilde{\Omega}_0(\mathbf{k}) = -\omega U_e\tilde{\alpha}''(\mathbf{k})\tilde{h}(\mathbf{k})$ denotes the 0$^{\text{th}}$-order component that includes the signals of electric dipole absorption, and $\tilde{\Omega}_{\pm 1}(\mathbf{k}) = C_0\tilde{G}''(\mathbf{k}\mp\mathbf{k}_C)\tilde{h}(\mathbf{k})$ denote the ±1$^{\text{st}}$-order components which contain only the high spatial frequency information on chiral domains. To separate the three components in Eq. (7), at least three subimages $M(\mathbf{r})$ with three individual pattern phases $\varphi$ are required. Each subimage in Fourier space $\tilde{M}_j(\mathbf{k})$ represents a linear equation that composes the three unknown components $\tilde{\Omega}_l(\mathbf{k} - l\mathbf{k}_C)$, where the subscript $j$ indicates the index of the subimages with the corresponding spatial phases of the structured illumination pattern. Once three subimages, namely three linear equations, are acquired by shifting the structured OC pattern twice (which correlates to three $\varphi_j$), the components $\tilde{\Omega}_l(\mathbf{k} - l\mathbf{k}_C)$ can be extracted individually by solving those linear equations. A preferable choice for the pattern phases $\varphi_j$ is a set of equidistant values between 0 and $2\pi$, i.e., $\varphi_j = (j-1)\frac{2\pi}{3} = (j-1)\varphi_0$, where $j = 1,2,3$. Eventually, this separation process can be expressed as the discrete Fourier transform of the three subimages

$$\tilde{\Omega}_l(\mathbf{k} - l\mathbf{k}_C) = -\frac{\varepsilon}{2\Phi_q \beta_d \eta \Delta t} \frac{1}{3} \sum_{j=1}^{3} \tilde{M}_j(\mathbf{k}) e^{-il(j-1)\varphi_0}, \qquad (8)$$

where the coefficient of 1/3 is a normalization factor. As noted in the previous section, the signals from the electric dipole absorption should be separated from that of CD responses. Because $\tilde{\Omega}_0(\mathbf{k})$ includes the contribution from the electric dipole absorption, we discard it after carrying out the component separation in order to obtain images of chiral domains. It is important to emphasize here that, although the emission coming from the electric dipole absorption is discarded, it still plays an important role in the signal-to-noise ratio analysis as the noise is a function of the overall emission intensity. To achieve isotropic resolution improvement (i.e., extend the effective OTF evenly in Fourier space), the illumination patterns with different in-plane orientations $\mathbf{k}_{C,\theta}$, which is indicated by the additional index $\theta$, are applied during the image acquisition. In this way, one can obtain the separated components $\tilde{\Omega}_{l,\theta}(\mathbf{k} - l\mathbf{k}_{C,\theta})$ situated at $\mathbf{k} = l\mathbf{k}_{C,\theta}$ in the Fourier space. Practically, the choice of the pattern orientations is $\theta = \{0°, 60°, 120°\}$ such that totally nine subimages (three pattern phases times three pattern orientations) are acquired. Next, those separated components $\tilde{\Omega}_{l,\theta}(\mathbf{k} - l\mathbf{k}_{C,\theta})$ are translated by $\pm\mathbf{k}_{C,\theta}$, i.e., $\tilde{\Omega}_{l,\theta}(\mathbf{k} - l\mathbf{k}_{C,\theta}) \to \tilde{\Omega}_{l,\theta}(\mathbf{k})$, to the correct position in the Fourier space for the following recombination. In the recombination process, $\tilde{\Omega}_{l,\theta}(\mathbf{k})$ are combined by applying the Wiener filter to yield the final reconstructed image

$$\tilde{M}_{rec}(\mathbf{k}) = \frac{\sum_{l=\pm 1,\theta} \tilde{h}^*(\mathbf{k} + l\mathbf{k}_{C,\theta}) \tilde{\Omega}_{l,\theta}(\mathbf{k})}{\sum_{l=\pm 1,\theta} \left|\tilde{h}(\mathbf{k} + l\mathbf{k}_{C,\theta})\right|^2 + w}, \qquad (9)$$

that contains high spatial frequency information on the chiral domains in the Fourier space. Here, $w$ denotes the empirical Wiener parameter. Note that $l = \pm 1$ because $\tilde{\Omega}_0(\mathbf{k})$ has been discarded previously to remove the contribution from the electric dipole absorption, i.e., the unmodulated fluorescence contribution. The reconstructed chiral SIM image in real space is obtained by performing the inverse Fourier transform of Eq. (9)

$$M_{rec}(\mathbf{r}) = \mathcal{F}^{-1}\left[\frac{\sum_{l=\pm 1,\theta} \tilde{h}^*(\mathbf{k}+l\mathbf{k}_{C,\theta})\tilde{\Omega}_{l,\theta}(\mathbf{k})}{\sum_{l=\pm 1,\theta}\left|\tilde{h}(\mathbf{k}+l\mathbf{k}_{C,\theta})\right|^2 + w}\right], \tag{10}$$

where $\mathcal{F}^{-1}$ denotes the operation of inverse Fourier transform. The whole process of the image reconstruction in chiral SIM is schematically summarized in Fig. 2.

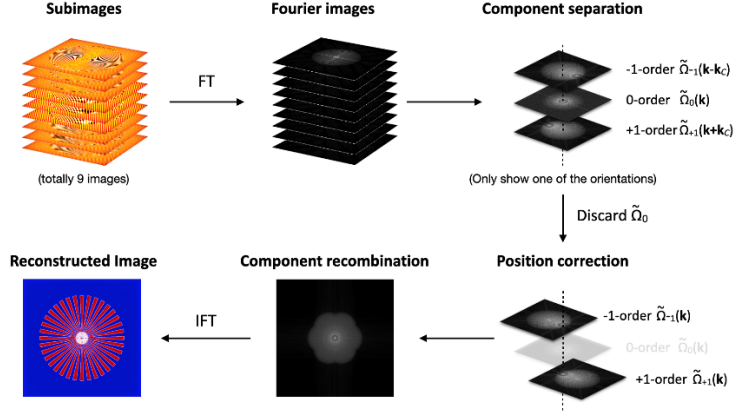

Fig. 2. Schematic illustration of the image reconstruction in chiral SIM. The magnitude of the spectrum in the Fourier images is enhanced for clarity.

### 3. The Effect of Noise in Chiral SIM

In a real experiment, the recorded subimages can be degraded by a variety of noises, such as the readout noise and the dark noise due to the readout process (e.g., charge transfer, analog-to-digital conversion, etc.) and thermally-induced free electrons. Besides, photons that arrive at the camera sensor show randomness, which inevitably leads to photon shot noise. In general, the readout noise of a modern imaging device, such as a scientific CMOS (sCMOS) camera, has been much reduced (~ 1 e$^-$ at high readout rate [27]) thanks to the advances of in-sensor architecture and fabrication. The dark noise can also be appropriately managed by cooling the camera (~ 0.1 e$^-$/pixel/s). Nevertheless, the photon shot noise cannot be eliminated by any practical means and is usually dominant among all sources of noise during the image acquisition. In this regard, we focus on the influence of the shot noise on the reconstructed chiral SIM image and omit the effects from the readout noise and dark noise as they are in principle addressable in the experiment.

*3.1 SNR of the reconstructed chiral SIM image*

Considering the photon shot noise $N_{shot}(\mathbf{r})$ that occurs during the image acquisition, the subimages in real space can be expressed as

$$M'_j(\mathbf{r}) = M_j(\mathbf{r}) + N_{shot,j}(\mathbf{r}), \tag{11}$$

where $M_j(\mathbf{r})$ denotes noiseless subimage as described by Eq. (5). Because all operations in the image reconstruction are, as illustrated in the previous section, performed in the Fourier space, we also consider the Fourier transform of Eq. (11)

$$\tilde{M}'_j(\mathbf{k}) = \tilde{M}_j(\mathbf{k}) + \tilde{N}_{shot,j}(\mathbf{k})$$

$$= -\frac{2\Phi_q \beta_d \eta \Delta t}{\varepsilon} \sum_{l=-1}^{1} \tilde{\Omega}'_l(\mathbf{k} - l\mathbf{k}_C) e^{il(j-1)\varphi_0}, \tag{12}$$

where $\tilde{\Omega}'_l(\mathbf{k} - l\mathbf{k}_C) = \tilde{\Omega}_l(\mathbf{k} - l\mathbf{k}_C) + \tilde{\chi}_l(\mathbf{k} - l\mathbf{k}_C)$ denotes the degraded separated components with the noise term

$$\tilde{\chi}_l(\mathbf{k} - l\mathbf{k}_C) = \frac{\varepsilon}{2\Phi_q \beta_d \eta \Delta t} \frac{1}{3} \sum_{j=1}^{3} \tilde{N}_{shot,j}(\mathbf{k}) e^{-il(j-1)\varphi_0}. \tag{13}$$

Following a derivation similar to that in Section 2.2, the reconstructed chiral SIM image in the presence of the shot noise becomes

$$\tilde{M}'_{rec}(\mathbf{k}) = \frac{\sum_{l=\pm 1,\theta} \tilde{h}^*(\mathbf{k} + l\mathbf{k}_{C,\theta}) \tilde{\Omega}'_{l,\theta}(\mathbf{k})}{\sum_{l=\pm 1,\theta} \left|\tilde{h}(\mathbf{k} + l\mathbf{k}_{C,\theta})\right|^2 + w}$$

$$= \tilde{M}_{rec}(\mathbf{k}) + \frac{\sum_{l=\pm 1,\theta} \tilde{h}^*(\mathbf{k} + l\mathbf{k}_{C,\theta}) \tilde{\chi}_{l,\theta}(\mathbf{k})}{\sum_{l=\pm 1,\theta} \left|\tilde{h}(\mathbf{k} + l\mathbf{k}_{C,\theta})\right|^2 + w}. \tag{14}$$

To identify the noise in the image, we calculate the variance of $\tilde{M}'_{rec}(\mathbf{k})$

$$\mathrm{Var}\left[\tilde{M}'_{rec}(\mathbf{k})\right] = \mathrm{Var}\left[\frac{\sum_{l=\pm 1,\theta} \tilde{h}^*(\mathbf{k} + l\mathbf{k}_{C,\theta}) \tilde{\chi}_{l,\theta}(\mathbf{k})}{\sum_{l=\pm 1,\theta} \left|\tilde{h}(\mathbf{k} + l\mathbf{k}_{C,\theta})\right|^2 + w}\right]$$

$$= \frac{\sum_{l=\pm 1,\theta} \left|\tilde{h}^*(\mathbf{k} + l\mathbf{k}_{C,\theta})\right|^2}{(\sum_{l=\pm 1,\theta} \left|\tilde{h}(\mathbf{k} + l\mathbf{k}_{C,\theta})\right|^2 + w)^2} \mathrm{Var}\left[\tilde{\chi}_{l,\theta}(\mathbf{k})\right]. \tag{15}$$

According to the calculation of $\mathrm{Var}\left[\tilde{\chi}_{l,\theta}(\mathbf{k})\right]$ in Appendix 6.2, Eq. (15) can be simplified as

$$\mathrm{Var}\left[\tilde{M}'_{rec}(\mathbf{k})\right] = \frac{\sum_{l=\pm 1,\theta} \left|\tilde{h}^*(\mathbf{k} + l\mathbf{k}_{C,\theta})\right|^2}{(\sum_{l=\pm 1,\theta} \left|\tilde{h}(\mathbf{k} + l\mathbf{k}_{C,\theta})\right|^2 + w)^2} \left(\frac{\varepsilon \omega U_e \tilde{\alpha}''(0)\tilde{h}(0)}{6\Phi_q \beta_d \eta \Delta t}\right). \tag{16}$$

The SNR of the reconstructed chiral SIM image in Fourier space is thus defined by

$$\mathrm{SNR}(\mathbf{k}) = \frac{\tilde{M}_{rec}(\mathbf{k})}{\sqrt{\mathrm{Var}\left[\tilde{M}'_{rec}(\mathbf{k})\right]}}$$

$$= \frac{C_0 \tilde{G}''(\mathbf{k})}{\sqrt{\omega U_e \tilde{\alpha}''(0)}} \sqrt{\frac{6}{\varepsilon} \Phi_q \beta_d \eta \Delta t} \sqrt{\frac{\sum_{l=\pm 1,\theta} \left| \tilde{h}(\mathbf{k} + l\mathbf{k}_{C,\theta}) \right|^2}{\tilde{h}(0)}} . \qquad (17)$$

Considering Eqs. (3) and (4), Eq. (17) can be further simplified as

$$\text{SNR}(\mathbf{k}) = \frac{\tilde{G}''(\mathbf{k})}{\sqrt{\tilde{\alpha}''(0)}} \frac{\cos^2 a}{c} E_0 \sqrt{6\omega \Phi_q \beta_d \eta \Delta t} \sqrt{\frac{\sum_{l=\pm 1,\theta} \left| \tilde{h}(\mathbf{k} + l\mathbf{k}_{C,\theta}) \right|^2}{\tilde{h}(0)}}$$

$$= \frac{\tilde{G}''(\mathbf{k})}{\sqrt{\tilde{\alpha}''(0)}} \frac{\cos^2 a}{c^{3/2}} \sqrt{\frac{12 \omega I_0 \Phi_q \beta_d \eta \Delta t}{n\varepsilon_0}} \sqrt{\frac{\sum_{l=\pm 1,\theta} \left| \tilde{h}(\mathbf{k} + l\mathbf{k}_{C,\theta}) \right|^2}{\tilde{h}(0)}}, \qquad (18)$$

where $I_0$ denotes the laser intensity on the sample. $c$ and $\varepsilon_0$ are the speed of light and permittivity in vacuum, respectively.

Eq. (18) is instructive because it shows that the SNR of the reconstructed chiral SIM image is governed by multiple quantities. First, the strength of the CD responses plays an important role because the SNR is directly proportional to the Fourier spectrum of the molecular chiral polarizability $\tilde{G}''(\mathbf{k})$. Secondly, as $\tilde{\alpha}''(0)$ is in the denominator of Eq. (18), chiral domains with smaller $\tilde{\alpha}''(0)$ are expected to show better SNR if other parameters remain the same. This makes it easier for the chirality-dependent signals of the sample to stand out from the dipolar absorption ones. Thirdly, enhancing the laser intensity on the sample $I_0$ also improves the SNR. However, it is important to note that excessive laser intensity can also lead to unwanted photobleaching of the fluorophores. Therefore, the excitation power should be carefully adjusted. Fourthly, the incident angle of the *s*- and *p*-polarized light $a$ also affects the SNR because the amplitude of the OC modulation $C_0$ is proportional to $\cos a$. The SNR thus increases with increasing $\cos^2 a$. However, this is at the price of decreasing the spatial frequency of the OC patterns as $|\mathbf{k}_C|$ is proportional to $\sin a$. Therefore, there should be a balance between increasing SNR and gaining a better resolution enhancement. Finally, increasing the exposure time $\Delta t$, selecting the fluorophore with high quantum yield $\Phi_q$, and improving the detection efficiency of the imaging system $\beta_d$, as well as increasing the fluorophore number $\eta$ within a voxel can all enhance the SNR.

## 4. Numerical Demonstration of chiral SIM in the presence of noise

To validate the analytical derivation of the SNR and illustrate the effect of noise on the quality of the reconstructed chiral SIM image, in this section, we present a series of numerical demonstrations using a synthetic sample containing randomly distributed chiral filaments with both handedness and calculate the corresponding SNR as a measure of the image quality. The chiral filaments shown in Fig. 3 (a) are applied to mimic a biological sample in a real experiment [9]. Each filament comprises different chiral domains of which chirality is distributed unevenly. This feature contrasts with the regular shapes shown in conventional standard targets (e.g., Siemens star) and renders the synthetic sample more realistic. In the numerical demonstration, these chiral filaments are assumed to be stained by the fluorophores that have similar molecular and optical properties to Rhodamine 6G (R6G), which is one of the most widely used dyes in fluorescence microscopy. This means we assume that the fluorophore can be excited by 532 nm laser and its emission peak appears at a spectral range longer than 550 nm [28]. In addition, the quantum yield $\Phi_q$ and the imaginary part of the polarizability $\alpha''$ of the fluorophore should be close to that of R6G, which are 0.95 [28] and $6 \times 10^{-38}$ [29], respectively. After the labeling process, the fluorescent molecules are supposed to freely rotate and form the corresponding fluorescent chiral assembly after attaching to the chiral structure of the filaments. In this condition, the criteria of FDCD are assumed to be satisfied and the fluorophores emit CD-dependent fluorescence, i.e., the emission strength is dependent on the circular dichroism. In our demonstration, the dissymmetry factor [24]

$$g_{\mathrm{CPL}}(\mathbf{r}) = 2\frac{A_L(\mathbf{r}) - A_R(\mathbf{r})}{A_L(\mathbf{r}) + A_R(\mathbf{r})} = \frac{2C_{\mathrm{CPL}}G''(\mathbf{r})}{\omega U_{e,\mathrm{CPL}}\alpha''(\mathbf{r})}, \tag{19}$$

is employed to describe the chiral responses of local domains within the filaments. Here, the subscript "CPL" indicates that the dissymmetry factor is measured upon circularly polarized excitation, and $A_L$ ($A_R$) denotes the absorption of left (right) CPL. $U_{e,\mathrm{CPL}}$ is the time-averaged electric energy density of CPL. Figs. 3(a) and (b) display $g_{\mathrm{CPL}}(\mathbf{r})$ of the filaments used in the simulations and the distribution of $g_{\mathrm{CPL}}(\mathbf{r})$, respectively. According to Fig. 3(b), the mean value of $g_{\mathrm{CPL}}(\mathbf{r})$ is around zero and the standard deviation is about $1.2 \times 10^{-3}$, similar to that of an ordinary chiral sample ($g_{\mathrm{CPL}}$ at the scale of $10^{-3} \sim 10^{-4}$).

The illumination patterns featured by the spatial distribution of $U_e(\mathbf{r})$ and $C(\mathbf{r})$ are simulated using the finite-difference time-domain method (FDTD Solutions, Lumerical). This mimics the condition of a real experiment. Two plane waves are superimposed on the sample plane to generate the required uniform $U_e(\mathbf{r})$ and spatially structured $C(\mathbf{r})$. The excitation

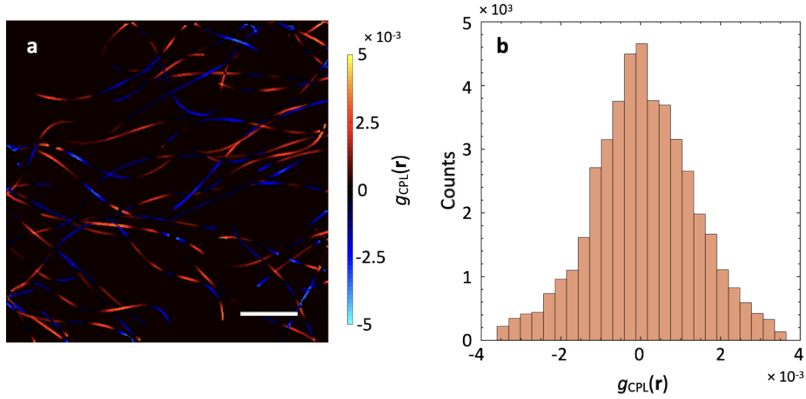

Fig. 3. (a) The local distribution of the dissymmetry factor $g_{\mathrm{CPL}}(\mathbf{r})$ of the synthetic randomly distributed chiral filaments used in the simulations. The color bar indicates the value of $g_{\mathrm{CPL}}(\mathbf{r})$. Scale bar: 2 μm. (b) The histogram showing the distribution of $g_{\mathrm{CPL}}(\mathbf{r})$ of the filaments in 3(a). The mean value is around zero and the standard deviation is about $1.2 \times 10^{-3}$.

wavelength is selected to be 532 nm, which overlaps with the absorption band of the targeted fluorophore. Because the emission wavelength of the fluorophores is around 550 nm and the numerical aperture (NA) of an oil immersion objective is selected to be 1.2, the theoretical cut-off frequency of the OTF $k_{\mathrm{cutoff}}$ is about $4.4 \times 10^{-3}$ nm$^{-1}$, as calculated from the reciprocal of

Abbe diffraction limit [30]. The pixel size in the numerical simulation is 20 × 20 nm² and we assume around five fluorophores reside in each pixel where the chiral filaments locate, i.e., $\eta$ = 5, which corresponds to a relatively high molecular density such that all chiral filaments are supposed to be fully-labeled and the distribution of $g_{\mathrm{CPL}}(\mathbf{r})$ can be well-characterized by the emitted fluorescence. The incident angle of two plane waves $a$ is set to be 41° to achieve a moderate resolution improvement of 1.85 (i.e., the ratio of $|\mathbf{k}_C|$ to $k_{\mathrm{cutoff}}$ is 0.85). By rotating the incident directions along the optical axis and adjusting the initial phases of the light sources, the structured OC patterns are generated in three orientations $\theta$ with three initial phases $\varphi_j$.

The formation of the subimages is computed according to Eq. (2) with the FDTD simulated illumination patterns in MATLAB R2017b (MathWorks, Natick, MA, USA) and the shot noise of the images is imposed using the noise function from the DIPimage toolbox [31]. The wide-field FDCD image using the Wiener deconvolution method [31] is also provided as a benchmark to highlight the resolution improvement of the reconstructed chiral SIM image. To stand out the noise effect with respect to the spatial frequency that relates to different features of the chiral domains, $\mathrm{SNR}(\mathbf{k})$ of the reconstructed chiral SIM image is averaged along the azimuthal angle in the Fourier domain and denoted by $\mathrm{SNR}_{\mathrm{avg}}(k)$ (see Appendix 6.3).

### 4.1 Effect of the strength of chiral responses on the chiral SIM image quality

Simulated chiral SIM images of a sample containing filaments with different dissymmetry factors are shown in Fig. 4. In order to view the impacts of the dissymmetry factor on the image quality, $g_{\mathrm{CPL}}(\mathbf{r})$ is multiplied by a factor of $m$, where $m$ = 1, 10 and 100. The exposure time $\Delta t$ and the laser intensity $I_0$ on the sample plane are set to be 1 s and 50 W/cm² for modest experiment conditions. All deconvolved wide-field FDCD and reconstructed chiral SIM images are normalized to their maximum pixel value within the region of interest hereafter. When $m$ = 1, lots of artifacts appear in both deconvolved wide-field FDCD image and chiral SIM image. These artifacts severely reduce the image quality, as can be seen in Figs. 4(a) and 4(d). This shows that the signal from samples with weak chiral response cannot compete with the shot noise and the information on the chiral domains is not restored properly by the deconvolution and the chiral SIM image reconstruction. In contrast, the image quality is apparently improved when $m$ = 10 and 100, at which the chiral response is not buried by the noise. This improvement can be clearly seen in the deconvolved wide-field FDCD and reconstructed chiral SIM images, as shown in Figs. 4(b), (c), (e), and (f). Besides, chiral SIM exhibits a superior resolving power compared to that of the conventional wide-field FDCD imaging method. The two filaments that are unresolved in Figs. 4(b) and (c) are well-separated in the chiral SIM images in Figs. 4(e) and (f), as marked by yellow arrows in Figs. 4(b), (c), (e), and (f). In general, the chiral SIM images shown in Figs. 4(d-f) clearly illustrate that the chiral SIM method requires the sample to have a sufficiently large dissymmetry factor. If the dissymmetry factor of the sample is too small (e.g., in Fig. 4(d)), the effect of noise becomes much pronounced than the signal modulation. This hampers the image reconstruction in chiral SIM. The noise is also unduly amplified in the high spatial frequency region, leading to the "hammer-stroke" artifacts after the deconvolution process [32]. In this case, one can hardly recognize the features of the filaments nor the enhanced resolution in the chiral SIM image.

Figure 5 shows the corresponding $\mathrm{SNR}_{\mathrm{avg}}(k)$ of the chiral SIM images shown in Figs. 4(d-f). As expected, the magnitude of $\mathrm{SNR}_{\mathrm{avg}}(k)$ at low spatial frequency is higher than that at high spatial frequency because the OTF of the collecting optics acts as a low-pass filter in the Fourier domain during the image acquisition. In the case of $m$ = 10 and 100, $\mathrm{SNR}_{\mathrm{avg}}(k)$ still remains sufficiently large even for $k > k_{cutoff}$. This agrees with the observation of the resolution improvement in the chiral SIM images displayed in Figs. 4(e) and (f). Eventually, $\text{SNR}_{avg}(k)$ drops drastically as $k$ approaches the theoretical bandwidth of the effective OTF, which is about 1.85-fold of $k_{cutoff}$. On the contrary, in the case of $m = 1$, $\text{SNR}_{avg}(k)$ is below unity when $k > k_{cutoff}$ and thus is supposed to be insufficient for a proper image reconstruction. This is confirmed as the reconstructed chiral SIM image shown in Fig. 4(d) is significantly degraded and cannot reveal the true features of the chiral domains.

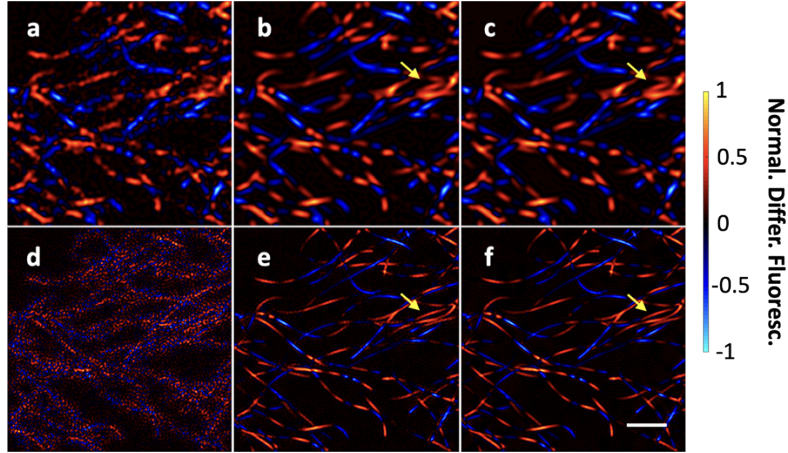

Fig. 4. Simulated wide-field FDCD images and chiral SIM images of chiral filaments with different dissymmetry factors controlled by the multiplication factor $m$. (a-c) Deconvolved wide-field FDCD images with $m = 1$, 10, and 100, respectively. (d-f) Reconstructed chiral SIM images with $m = 1$, 10, and 100, respectively. In this demonstration, $\Delta t = 1$ s and $I_0 = 50$ W/cm$^2$. The color bar indicates the value of the normalized differential fluorescence. Scale bar: 2 µm.

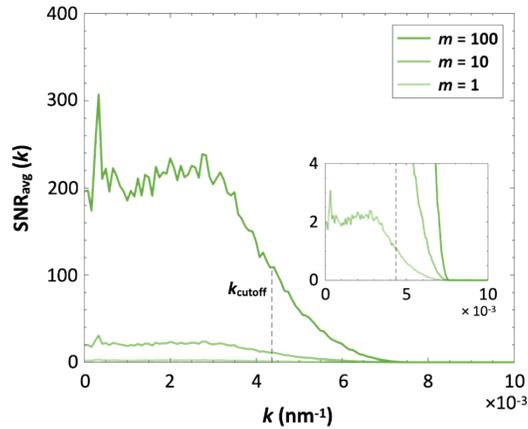

Fig. 5. $\text{SNR}_{avg}(k)$ when $m = 1$, 10, and 100. The black dashed line indicates $k_{cutoff}$ of the optical system. The inset shows the zoom-in of the green curves. In the numerical demonstration, $k_{cutoff} \approx 4.4 \times 10^{-3}$ nm$^{-1}$ when the emission wavelength of the fluorophore and NA of the objective are 550 nm and 1.2, respectively.

*4.2 Effect of the exposure time on the chiral SIM image quality*

Fig. 4(d) shows that the FDCD signals emitted from the chiral filaments with $m = 1$ are not sufficient to overcome the shot noise. As Eq. (18) suggests, prolonging $\Delta t$ also enhances the SNR. Figure 6 shows the $\mathrm{SNR}_{\mathrm{avg}}(k)$ at $k = 1.85\, k_{\mathrm{cutoff}}$ and $I_0 = 50$ W/cm² with respect to different $m$ and $\Delta t$. The ratio of $|\mathbf{k}_C|$ to $k_{\mathrm{cutoff}}$ is set to be 0.85 to gain a moderate resolution improvement. Therefore, the theoretical bandwidth of the effective OTF in the Fourier space is at $1.85\, k_{\mathrm{cutoff}}$ and the $\mathrm{SNR}_{\mathrm{avg}}(1.85\, k_{\mathrm{cutoff}})$ has to reach a certain value in order to obtain reasonable image quality for the visualization of resolution improvement. Since the image obtained with $m = 10$ and $\Delta t = 1$ s (Fig. 4(e)) demonstrates sufficiently good image quality, we use its corresponding $\mathrm{SNR}_{\mathrm{avg}}(1.85\, k_{\mathrm{cutoff}})$ as a threshold for comparison in Fig. 6 (indicated by a black arrow). Any experimental conditions that lead to $\mathrm{SNR}_{\mathrm{avg}}(1.85\, k_{\mathrm{cutoff}})$ in the grey zone in Fig. 6 cannot generate reasonably good images. For $\Delta t = 100$ s or larger, the $\mathrm{SNR}_{\mathrm{avg}}(1.85\, k_{\mathrm{cutoff}})$ would always be larger than the threshold (see the darkest red curve in Fig. 6). This conclusion is further verified by numerical simulations. The simulated chiral SIM images of the filament sample with $m = 1$ and $\Delta t = 10$ s, 50 s and 100 s are displayed in Figs. 7(d-f), respectively. $I_0$ has been set to 50 W/cm² for moderate illumination. Clearly, the image quality is enhanced as $\Delta t$ is increased. The chiral SIM image of $\Delta t = 100$ s shown in Fig. 7(f) has an image quality similar to that of the image in Fig. 4(e). For the image with $\Delta t = 10$ s in Fig. 7(d), although the resolution improvement can still be recognized, the sparse artifacts may cause impropriate interpretation of the image. Overall, the quality of the simulated chiral SIM images in Fig. 7 is in good agreement with the expectation inferred from the theoretical curves shown in Fig. 6.

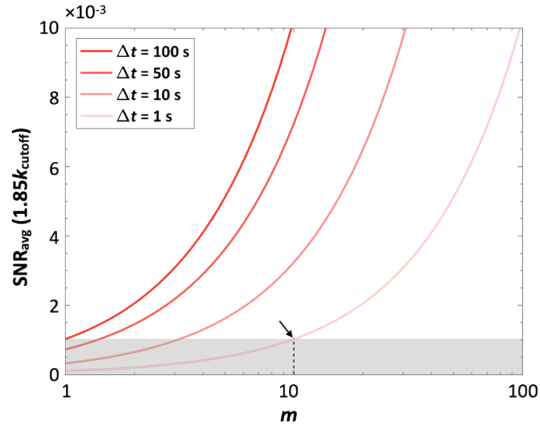

Fig. 6. $\mathrm{SNR}_{\mathrm{avg}}(1.85\, k_{\mathrm{cutoff}})$ as a function of $m$, plotted for $\Delta t = 1$ s, 10 s, 50 s, and 100 s. The black arrow indicates the corresponding $\mathrm{SNR}_{\mathrm{avg}}(1.85\, k_{\mathrm{cutoff}})$ when $m = 10$ and $\Delta t = 1$. The experimental conditions that lead to $\mathrm{SNR}_{\mathrm{avg}}(1.85\, k_{\mathrm{cutoff}})$ in the grey zone cannot generate chiral SIM images with good quality.

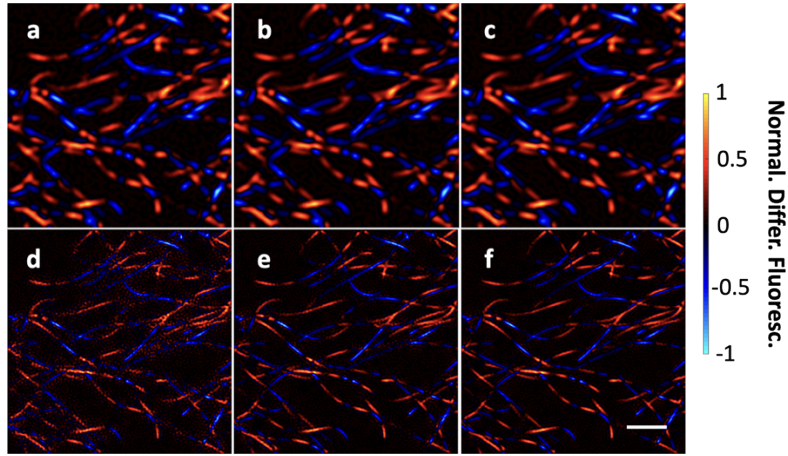

Fig. 7. Simulated wide-field FDCD images and chiral SIM images on a sample containing chiral filaments with $m = 1$ at different exposure time $\Delta t$. The illumination power is set to $I_0 = 50$ W/cm². (a-c) Deconvolved wide-field FDCD images with $\Delta t$ = 10 s, 50 s, and 100 s, respectively. (d-f) Reconstructed chiral SIM images with $\Delta t$ = 10 s, 50 s, and 100 s, respectively. The color bar indicates the value of the normalized differential fluorescence. Scale bar: 2 μm.

### *4.3 Effect of the laser intensity on the chiral SIM image quality*

Although prolonging $\Delta t$ can fairly increase the quality of the chiral SIM image, it is at the expense of increasing acquisition time, which is not preferred for rapid imaging. Alternatively, one can think about enhancing the SNR by increasing the excitation power $I_0$. However, increasing the excitation power is also unfavored in fluorescence microscopy as it may lead to photobleaching of the fluorophores. Moreover, as we will show in the following, increasing the laser power is not as effective in increasing SNR as the methods discussed in the previous sections. Figure 8 shows the curves of $\text{SNR}_{\text{avg}}(1.85\,k_{\text{cutoff}})$ as a function of $m$ under different excitation power $I_0$, ranging from 50 W/cm² to 5000 W/cm². This power range is comparable to the laser intensity applied in typical wide-field fluorescence microscopy. In a typical fast imaging process, the frame rate can be 5.5 fps or higher. However, for chiral SIM, the overall frame rate also depends on the time needed for the image reconstruction. Therefore, we set $\Delta t$ to be 20 ms for a reasonable frame rate of 5 fps. As can be seen in Fig. 8, for 1 < $m$ < 10, all the curves are in the grey zone. This means that even at the lowest frame rate and the high excitation power of 5000 W/cm², the FDCD signal is still insufficient to overcome the shot noise for the reconstruction of reasonably good images. This conclusion is confirmed by the simulated images shown in Fig. 9. The reconstructed chiral SIM images have been simulated with $\Delta t$ = 20 ms and $m$ = 1. Apparently, all deconvolved wide-field FDCD and reconstructed chiral SIM images in Fig. 9 are degraded by the noise. In particular, the artifacts in the chiral SIM images (Figs. 9(d-f)) are prevailing such that even the global shape of the filaments can hardly be identified. Although the resolution improvement could be vaguely seen in the chiral SIM image shown in Fig. 9(f), the image quality is still poor and therefore may lead to a flawed interpretation of the chirality distribution and relevant structural features.

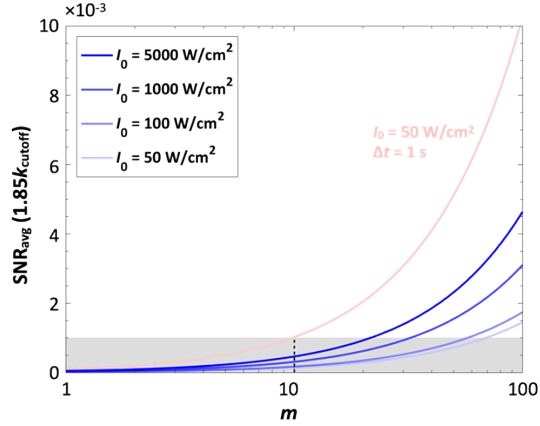

Fig. 8. $\text{SNR}_{\text{avg}}(1.85\,k_{\text{cutoff}})$ as a function of $m$, plotted when $I_0 = 50$ W/cm², 100 W/cm², 1000 W/cm², and 5000 W/cm². $\Delta t$ is set to be 20 ms for fast imaging the chiral sample. The light red curve indicates the $\text{SNR}_{\text{avg}}(1.85\,k_{\text{cutoff}})$ when $I_0 = 50$ W/cm² and $\Delta t = 1$ s, as shown in Fig. 6. The experimental condition (e.g., $I_0 = 50$ W/cm² and $m = 10$) of which $\text{SNR}_{\text{avg}}(1.85\,k_{\text{cutoff}})$ is within the grey zone could be considered as incapable of acquiring a decent final chiral SIM image.

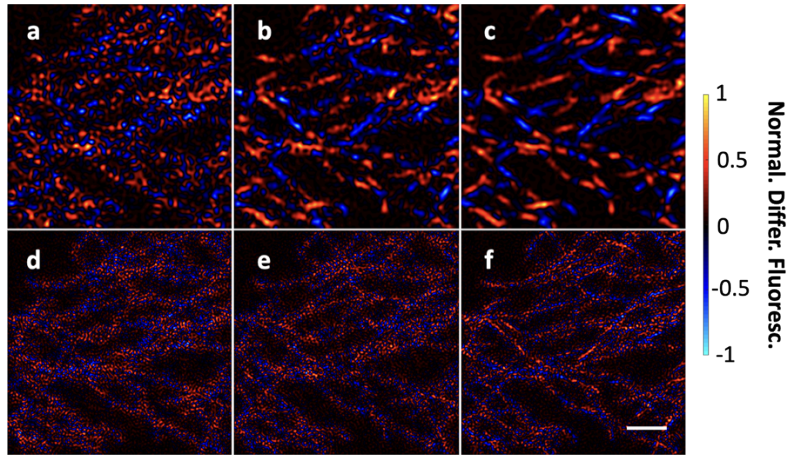

Fig. 9. Simulated images of wide-field FDCD images and chiral SIM images with respect to different excitation intensities $I_0$. (a-c) Deconvolved wide-field FDCD images with $I_0 = 100$ W/cm², 1000 W/cm², and 5000 W/cm², respectively. (c-f) Reconstructed chiral SIM images with $I_0 = 100$ W/cm², 1000 W/cm² and 5000 W/cm², respectively. In this demonstration, $m = 1$ and $\Delta t = 20$ ms. The color bar indicates the value of the normalized differential fluorescence. Scale bar: 2 μm.

## 5. Conclusion

In summary, we have illustrated the principle of image formation and image reconstruction in chiral SIM and analytically derived the SNR of the chiral SIM image in the Fourier space. The spatial frequency-dependent SNR, i.e., $\text{SNR}(\mathbf{k})$, is used to investigate and optimize the performance of the proposed chiral SIM method in various experimental conditions. The influences of shot noise on the reconstructed chiral SIM image have been identified through the different numerical demonstrations. Approaches to improve the SNR of the reconstructed chiral SIM image, such as prolonging the exposure time and increasing laser intensity, are demonstrated. The quality of chiral SIM images significantly varies under different experimental conditions. First, the demonstration shown in Fig. 4 emphasizes the importance of chiral responses on the final image quality. When the dissymmetry factor of the chiral sample is in the range of $10^{-2} \sim 10^{-1}$, the FDCD signal is sufficient for performing chiral SIM with an exposure time of at least 1 s for each subimage. For samples whose local dissymmetry factor is smaller than $10^{-3}$, prolonging exposure time may be a practical way to improve the image quality. When the exposure time of each subimage extends up to 100 s, chiral SIM is able to restore the high spatial information on the chiral domains with decent quality even though the dissymmetry factor of the sample is as low as $10^{-3} \sim 10^{-4}$, as illustrated in Fig. 7(f). However, long exposure time limits the imaging speed. Increasing the laser power only offers a small enhancement in the image quality. But one should keep in mind that excess laser intensity may induce unwanted photobleaching of the fluorophores. There should be a trade-off between the photobleaching and imaging frame rate. Thus, fast imaging for chiral samples with the dissymmetry factor smaller than $10^{-3}$ remains very difficult for the proposed chiral SIM method. Innovative illumination schemes, such as chiral Bloch surface waves [33,34] or plasmonic [35-37] and dielectric [38-41] nanostructures, might be promising solutions to these issues as they offer the possibility to manipulate the optical field such that the weak chiral light-matter interactions can be enhanced. The main challenge to utilize these illumination schemes is how to engineer the electric energy density and OC distribution of the illumination fields in order to satisfy the requirements of chiral SIM. This is particularly challenging for nanostructures, where the local electromagnetic field is primarily concentrated in the proximity of the structures and thus difficult to be tailored and shifted laterally, as required in the proposed chiral SIM method. The spin-orbit coupling also makes it difficult to maintain the spatial distribution of the local electric energy density while flipping the handedness of OC. Further efforts in this direction are necessary for the application of these advanced illumination schemes for chiral SIM.

## 6. Appendix

### 6.1 Eliminating the fluorescence contribution from electric dipole absorption by image subtraction

Considering each pattern phase, $U_e(\mathbf{r})$ should remain the same when the structured $C(\mathbf{r})$ changes its handedness or distribution in order to eliminate the unmodulated fluorescence contribution from the electric dipole absorption by image subtraction. In this way, we can express the conjugate images $M_{+,j}(\mathbf{r})$ and $M_{-,j}(\mathbf{r})$ as

$$M_{\pm,j}(\mathbf{r}) = \frac{2\Phi_q \beta_d \eta \Delta t}{\varepsilon} \Big[ \omega U_{e,j}(\mathbf{r}) \alpha''(\mathbf{r}) - C_{\pm,j}(\mathbf{r}) G''(\mathbf{r}) \Big] \otimes h(\mathbf{r}), \qquad (20)$$

where $j$ is the index of the spatial phase of the illumination pattern. When one conjugate image is subtracted from another, the resulting image is given by

$$\begin{aligned}\Delta M_j(\mathbf{r}) &= M_{+,j}(\mathbf{r}) - M_{-,j}(\mathbf{r}) \\ &= -\frac{2\Phi_q \beta_d \eta \Delta t}{\varepsilon} \Delta C_j(\mathbf{r}) G''(\mathbf{r}) \otimes h(\mathbf{r}),\end{aligned} \qquad (21)$$

where $\Delta C_j(\mathbf{r}) = C_{+,j}(\mathbf{r}) - C_{-,j}(\mathbf{r})$ is the differential OC distribution. In Eq. (21), the contribution from the electric dipole absorption is eliminated. Thus, as long as $\Delta C_j(\mathbf{r})$ is still in a cosinusoidal form, $\Delta M_j(\mathbf{r})$ can serve as the subimage and be applied in the chiral SIM image reconstruction.

*6.2 Derivation of* $\mathrm{Var}\left[\tilde{\chi}_{l,\theta}(\mathbf{k})\right]$

$$\begin{aligned}
\mathrm{Var}\left[\tilde{\chi}_{l,\theta}(\mathbf{k})\right] &= \mathrm{Var}\left[\frac{\varepsilon}{2\Phi_q \beta_d \eta \Delta t} \frac{1}{3}\sum_{j=1}^{3}\tilde{N}_{shot,j}(\mathbf{k})\right] \\
&= \left(\frac{\varepsilon}{6\Phi_q \beta_d \eta \Delta t}\right)^2 \mathrm{Var}\left[\sum_{j=1}^{3}\sum_{\mathbf{r}}N_{shot,j}(\mathbf{r})\right] \\
&= \left(\frac{\varepsilon}{6\Phi_q \beta_d \eta \Delta t}\right)^2 \sum_{j=1}^{3}\sum_{\mathbf{r}}M_j(\mathbf{r}) \\
&= \left(\frac{\varepsilon}{6\Phi_q \beta_d \eta \Delta t}\right)^2 \times \frac{6\Phi_q \beta_d \eta \Delta t}{\varepsilon}\tilde{\Omega}_0(0) \\
&= \frac{\varepsilon \omega U_e \tilde{\alpha}''(0)\tilde{h}(0)}{6\Phi_q \beta_d \eta \Delta t}.
\end{aligned} \qquad (22)$$

*6.3 Azimuthally-averaged SNR in the Fourier domain*

The averaged $\mathrm{SNR}(\mathbf{k})$ along the azimuthal angle in the Fourier plane of $\mathbf{k}_x$ and $\mathbf{k}_y$, can be expressed as

$$\mathrm{SNR}_{avg}(k) = \frac{\sum_{|\mathbf{k}|=k}\mathrm{SNR}(\mathbf{k})}{Q}, \qquad (23)$$

to demonstrate the SNR of the final image with respect to different spatial frequencies of the chiral filaments, regardless of the orientations in the Fourier space. Here, $Q$ denotes the number of the discrete data points in the Fourier space as $|\mathbf{k}| = k$.

*6.4 Summary of the parameters applied in the chiral SIM numerical demonstrations*

Table 1 sumarizes the parameters that are applied in the numerical demonstrations.

Table 1. Parametes used in the numerical demonstration

| Symbol | Name | Value |
|---|---|---|
| - | Pixel size | 20 nm |
| - | Excitation wavelength | 532 nm |
| - | Emission wavelength | 550 nm |
| $a$ | Incident angle | 41° |
| $I_0$ | Laser intensity on sample | 50 ~ 5000 W/cm² |
| $\Delta t$ | Exposure time of each subimage | 20 ms ~ 100 s |
| $\alpha''$ | Imaginary part of polarizability | $6 \times 10^{-38}$ (S.I. unit) |
| $\eta$ | Number of fluorophores in a voxel | 5 |
| $\Phi_q$ | Quantum yield | 0.95 |
| $\beta_d$ | Detection efficiency | 0.85 |
| NA | Numerical aperture | 1.2 |
| $n$ | Refractive index | 1.518 |

**Funding**

IPHT Innovation Project 2017, DFG SFB 1375 (C1 sub-project), DFG HU2626/3-1 (423427290), DFG 2626/5-1(445415315), and DFG 2626/6-1(447515653).

**Acknowledgements**

We acknowledge Prof. Rainer Heintzmann for providing the SIM reconstruction program and also thank Prof. Jiwei Zhang for the fruitful discussions.

**Disclosures**

The authors declare the following competing financial interest(s): J.-S.H. has filed a patent application for chiral SIM.